\documentclass{elsart}

\usepackage{natbib}

\usepackage{epsfig}

\usepackage{amssymb}
\usepackage{color}

\begin{document}
%
\def\etal{et  al.\ }
\def\araa{{Ann.\ Rev.\ Astron.\ Ap.}}
\def\aplet{{Ap.\ Letters}}
\def\aj{{Astron.\ J.}}
\def\apj{ApJ}
\def\apjl{{ApJ\ (Lett.)}}
\def\apjs{{ApJ\ Suppl.}}
\def\aas{{Astron.\ Astrophys.\ Suppl.}}
\def\aa{{A\&A}}
\def\aap{{A\&A}}
\def\mnras{{MNRAS}}
\def\nat{{Nature}}
\def\pasa{{Proc.\ Astr.\ Soc.\ Aust.}}
\def\pasp{{P.\ A.\ S.\ P.}}
\def\pasj{{PASJ}}
\def\pre{{Preprint}}
\def\sovlet{{Sov. Astron. Lett.}}
\def\adspr{{Adv. Space. Res.}}
\def\expas{{Experimental Astron.}}
\def\ssr{{Space Sci. Rev.}}
\def\apss{{Astrophys. and Space Sci.}}
\def\inpress{in press.}
\def\souspresse{sous presse.}
\def\inprep{in preparation.}
\def\enprep{en pr\'eparation.}
\def\submit{submitted.}
\def\soumis{soumis.}
\def\aph{{Astro-ph}}
\def\astroph{{Astro-ph}}

\begin{frontmatter}



\title{Simulating the impact of dust cooling on the statistical properties of the intracluster medium}


\author{Etienne Pointecouteau}
\address{Universit\'e de Toulouse/CNRS, CESR, 9 av. du Colonel Roche, 31400 Toulouse, France}
\ead{pointeco@cesr.fr}

\author{Antonio da Silva}
\address{CAUP, Rua das Estrelas, 4150-762 Porto, Portugal}
\address{Institut d'Astrophysique Spatiale, Bat 121, Universite Paris Sud, 91405 Orsay, France}

\author{Andrea Catalano}
\address{Université Paris Diderot, APC,  10, rue Alice Domon et Léonie Duquet, 75205 PARIS cedex 13, France}

\author{Ludovic Montier, Joseph Lanoux, Mauro Roncarelli, Martin Giard} 
\address{Universit\'e de Toulouse/CNRS, CESR, 9 av. du Colonel Roche, 31400 Toulouse, France}

\begin{abstract}
From the first stages of star and galaxy formation, non-gravitational
processes such as ram pressure stripping, SNs, galactic winds, AGNs,
galaxy-galaxy mergers, etc... lead to the enrichment of the IGM in
stars, metals as well as dust, via the ejection of galactic material
into the IGM.  We know now that these processes shape, side by side
with gravitation, the formation and the evolution of structures.  We
present here hydrodynamic simulations of structure formation
implementing the effect of the cooling by dust on large scale
structure formation. We focus on the scale of galaxy clusters and
study the statistical properties of clusters. Here we present our
results on the $T_X-M$ and the $L_X-M$ scaling relations which exhibit
changes on both the slope and normalization when adding cooling by
dust to the standard radiative cooling model. For example, the
normalization of the $T_X-M$ relation changes only by a maximum of 2\%
at $M=10^{14}$~M$_\odot$ whereas the normalization of the $L_X-T_X$
changes by as much as 10\% at $T_X=1$~keV for models that including
dust cooling. Our study shows that the dust is an added
non-gravitational process that contributes shaping the thermodynamical
state of the hot ICM gas.
\end{abstract}

\end{frontmatter}

\hspace*{-1cm}
\section{Introduction}
~\\[-3em] \indent The high metallicity observed in the intra-cluster
medium (ICM hereafter -- see for instance \citet{borgani08}, for
recent works) is understood within analytical models and numerical
simulations as a consequence of the various gravitational
(galaxy-galaxy mergers, galaxy interactions, ram pressure stripping)
and non-gravitational (SN powered galactic winds, AGN outbursts,
starburts winds) processes at play in the ICM. These processes drive
the formation and the evolution of galaxies within their environment
(see for instance works by
\citet{kapferer06,domainko06,moll07}. {Moreover, it is obvious that
the process of tearing material from galaxies does not discriminate between metals, gas, stars and dust. All are injected in the ICM/IGM further on leading to the metal enrichment of the surrounding medium, and possibly to its dust enrichment}.
To date the searches for a diffuse dust component within the ICM have
not been significantly successful. Together with their study of the
reddening of background galaxies towards clusters, \citet{muller08}
gathered the existing constraints on the presence of dust in the ICM
and the derived constraints in terms of extinction or dust-to-gas mass
ratio.
\citet{montier05} obtained a {clear statistical
  detection for the total dust emission in clusters (from member
  galaxies and ICM) by using the stacking of the} IRAS data (at 60 and
100~$\mu$m near the peak of dust emission) at the location of more
than 11000 clusters and groups. In an extension of this work,
\citet{giard08} have investigated for the first time the evolution of
the { total IR bolometric luminosity of clusters} with
redshift. Comparing to the expected IR luminosity from the integrated
SFR in cluster galaxies and considering the low end of current
constraints for the dust-to-gas mass ratio in the ICM
\citep{chelouche07}, 
{ \citet{giard08} concluded that ICM dust could
represent 2\% of the bolometric IR luminosity of clusters.}
On the other hand, from a theoretical point of view the effect of dust
on a ICM/IGM-type thermalized plasma has been formalised by
\citet{montier04}. These authors have computed the cooling function of
dust within an optically thin hot thermal plasma taking into account
the energetic budget for dust.
In view of their results, dust thus comes within the ICM/IGM, as an
added  {non-gravitational process} that may influence the
formation and evolution of large scale structures in a non-negligible
way.

In order to address this problem, we have put into place the first
hydrodynamic $N$-body numerical simulations of hierarchical structure
formation implementing the effect of dust cooling according to the
dust nature and abundance. In this proceedings paper, we present
original preliminary results of this study focussing on the properties
of clusters of galaxies at redshift zero.

\begin{figure}
\label{fig:lamdust}
\begin{center}
\hspace*{-0.5cm}\includegraphics*[height=6.5cm]{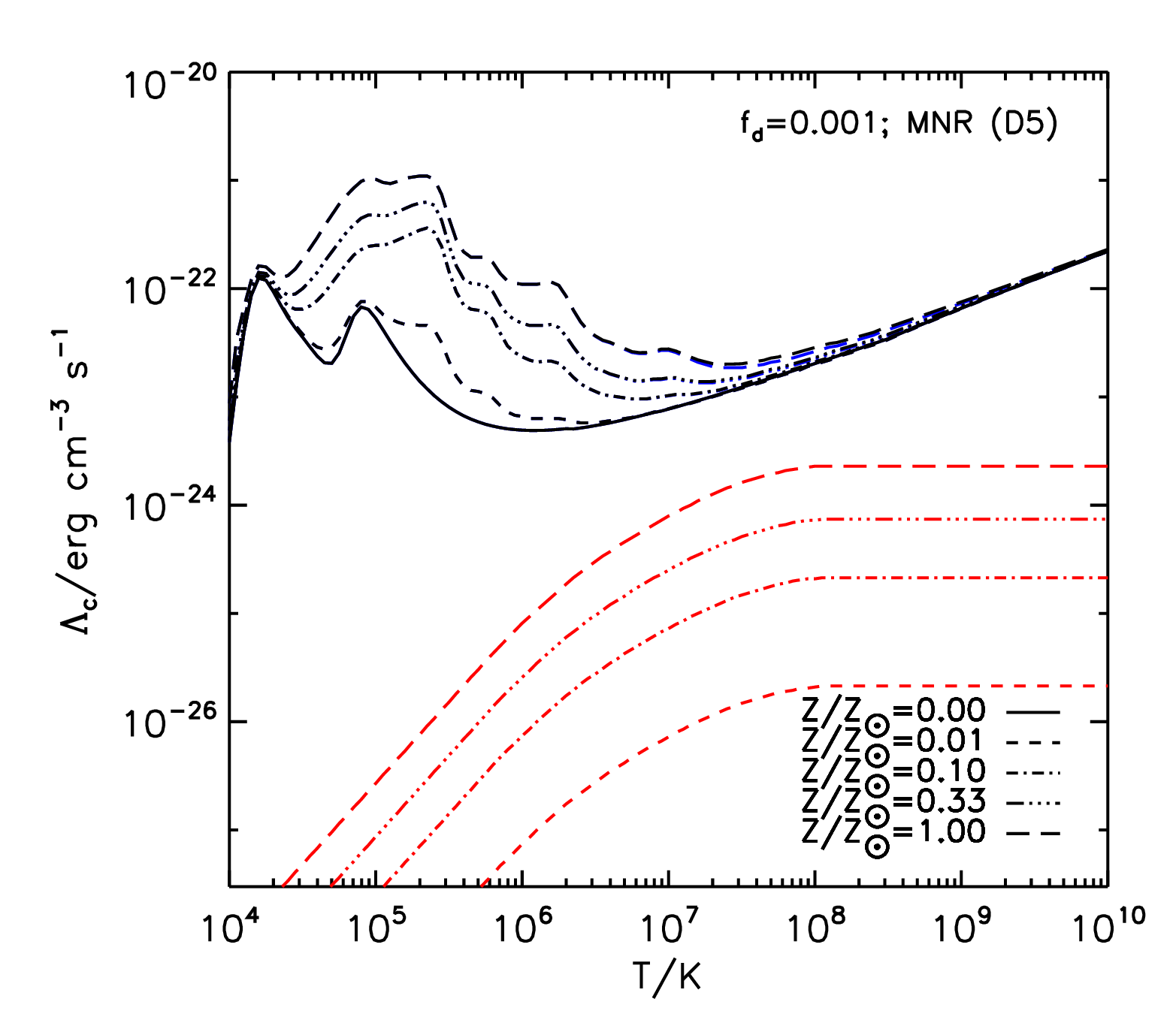}%
\includegraphics*[height=6.5cm]{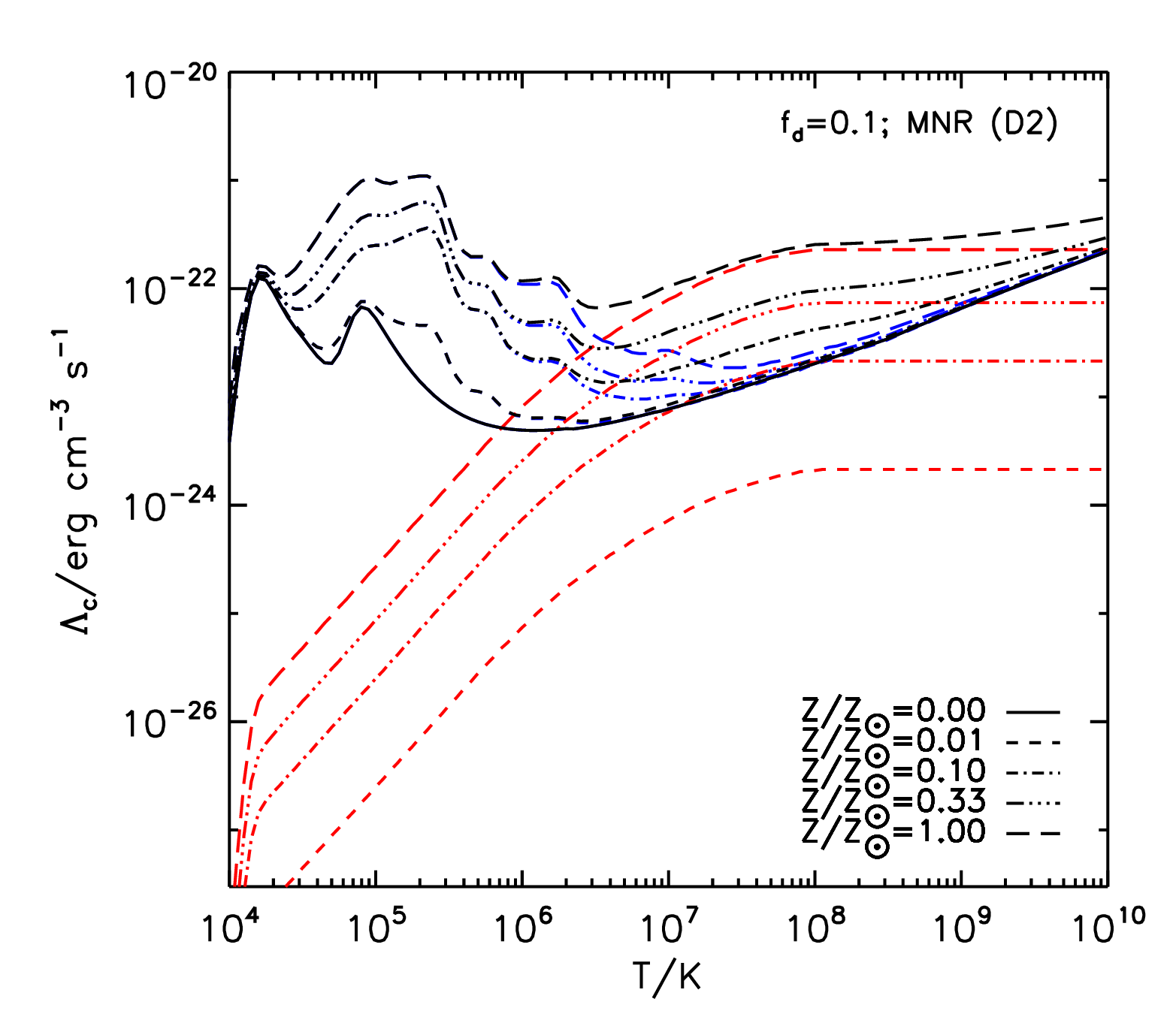}%
\end{center}
\caption{ Total cooling (radiative + dust) functions implemented in
  the numerical simulations (black lines).  The radiative and dust
  cooling functions alone are respectively shown in blue and red. The
  figure illustrate the dependency of the dust cooling efficiency as a
  function of the metallicity if the IGM (red curves on both panels)
  and as a function of the dust-to-gas mass ratio within the ICM (left
  and right panels).}
\end{figure}

\section{Cooling the ICM with dust}
\label{sec:cool}
~\\[-3em]
\indent \citet{montier04} have shown that in the temperature range of
galaxy clusters, dust can be a cooling agent of the ICM, which can
compete efficiently with radiative cooling. The strength of cooling by
dust depends on dust grain properties and more specifically on dust
abundance and dust grain sizes. The more abundant the dust and the
smaller the grains, the higher the efficiency of the cooling by dust
is.
\vspace*{-1em}
\begin{itemize}
\item The dust abundance, i.e the dust-to-gas mass ratio, is a
  function of the relative abundance of dust with respect to the solar
  vicinity dust abundance (i.e $Z_{d, \odot}$) and of the metallicity
  of the ICM (i.e $Z/Z_\odot$) : $Z_d=M_{dust}/M_{gas}=f_d\;
  (Z/Z_{\odot})\; Z_{d, \odot}$.  We investigate three values for
  $f_d$: $0.001$, $0.01$ and $0.1$. The corresponding dust-to-gas mass
  ratios bracket roughly the current theoretical and observational
  constraints on dust abundance in the ICM/IGM, i.e $10.^{-5}$ and
  $10^{-3}$ \citep{popescu00,aguirre01,chelouche07,muller08,giard08}.
\item At a given metallicity, the most relevant parameter for the
  cooling function by dust is the grain size. Indeed, the smaller the
  grains, the higher the cooling power of the dust is. We chose to test
  three types of size distribution: two fixed grain sizes with
  $a=10^{-3}$~$\mu$m and $a=0.5$~$\mu$m, respectively labelled {\it
    small} and {\it big}. The third assumes for the IGM dust grains a
  distribution in sizes as defined by \citet{mathis77} for the
  galactic dust: {$N(a)\propto a^{-3.5}$} within the size interval of
  $[0.001,0.5]$~$\mu$m. It is further referred as the `MNR'
  distribution.
\end{itemize}   
\vspace*{-1em}
Thus in our simulations, once the metallicity is known, $a$ and $f_d$
are the only two parameters driving the dust cooling rate (i.e
$\Lambda(a,Z)=\Lambda(a,f_d)$). Fig.~1 illustrates the
dependency of the dust cooling function with respect to the
metallicity and dust abundances for a MNR dust grain size
distribution.

\begin{figure}
\label{fig:ps}
\includegraphics*[width=\textwidth]{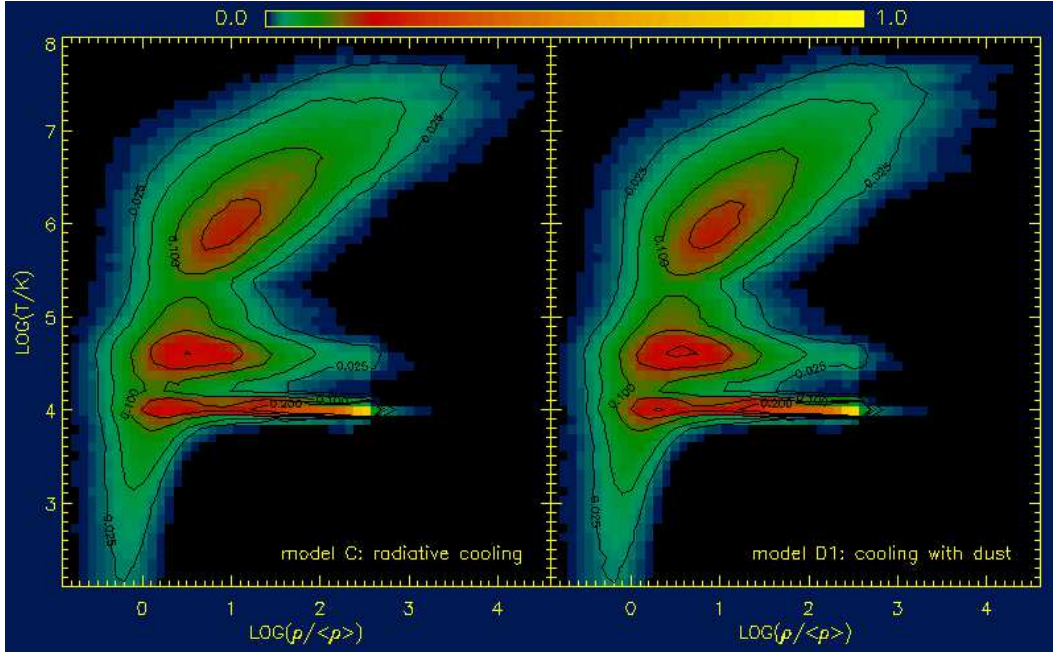}%
\caption{Density-Temperature phase space diagram of the radiative
  cooling simulation (left) and a case of radiative+dust cooling
  simulation. The separation of the different phases (ICM, IGM, cool
  gas) is illustrated by the dashed-yellow lines. {The
    overlaid contours are isoprobability contours of 0.025, 0.05, 0.1,
    0.15 and 0.2.}}
\end{figure}
\section{Simulation of structure formation with dust cooling}
\subsection{Hydrodynamic $N$-body simulations}
\label{sec:simu}
~\\[-3em]
\indent We have run a set of seven numerical simulations using the
public code package {\tt Hydra}, an adaptive
particle-particle/particle-mesh (AP$^3$M), gravity solver with
smoothed particle hydrodynamics (SPH)
\citep{couchman:1995,couchman:1991,pearce:1997}.  The global gas
metallicity is assumed to evolve with time as $Z=0.3(t/t_0)Z_\odot$,
where $Z_\odot$ is the solar metallicity and $t/t_0$ is the age of the
universe in units of the present time.  All simulations are issued
from the same initial condition snapshot, at $z=49$.  The initial
density field was constructed, using $N=4,096,000$ particles of
baryonic and dark matter, perturbed from a regular grid of fixed
comoving size $L=100 \, h^{-1} {\rm Mpc}$.  We assumed a $\Lambda$-CDM
cosmology with parameters, $\Omega =0.3$, $\Omega_{\Lambda}=0.7$,
$\Omega_{b}=0.0486$, $h=0.7$. The amplitude of the
matter power spectrum was normalised using $\sigma_8=0.9$.  With this
choice of parameters, the dark matter and baryon particle masses are
$2.1\times 10^{10} \, h^{-1} {\rm M_{\odot}}$ and $2.6 \times 10^{9}
\, h^{-1} {\rm M_{\odot}}$ respectively. The gravitational softening
in physical coordinates was $25\,h^{-1} {\rm kpc}$ below $z=1$ and
above this redshift scaled as $50(1+z)^{-1}\,h^{-1} {\rm kpc}$.
Gas in the simulations is allowed to cool using the total table
presented in Fig~1.  At a given time step, gas particles with
overdensities (relative to the critical density) larger than $10^4$,
and temperatures below $1.2\times 10^4$K are converted into
collisionless baryonic matter and no longer participate in the gas
dynamical processes.
\begin{figure}
\label{fig:maps}
\begin{center}
\includegraphics*[width=0.45\textwidth]{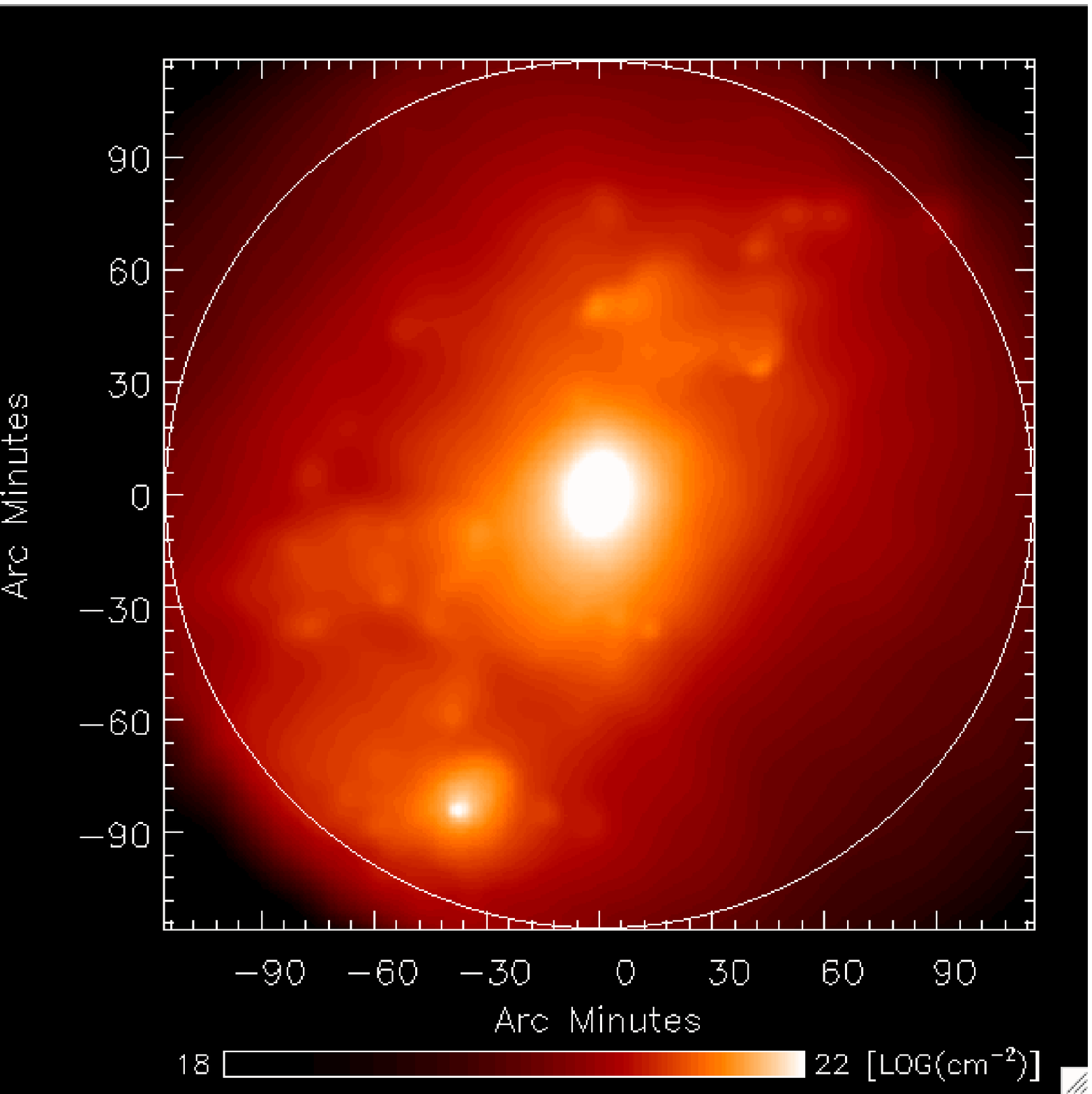}%
\includegraphics*[width=0.45\textwidth]{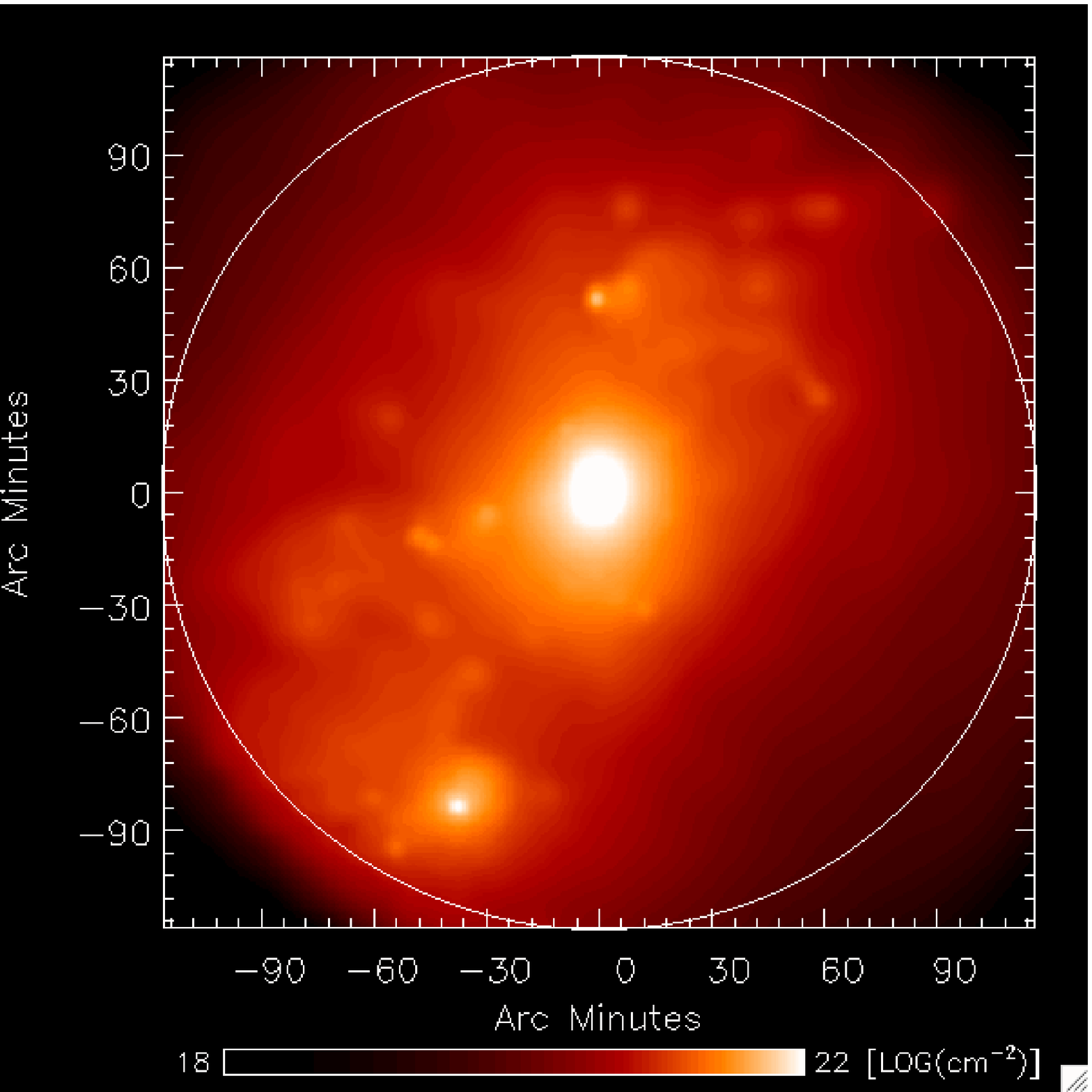}\vspace*{-0.2ex}
\includegraphics*[width=0.45\textwidth]{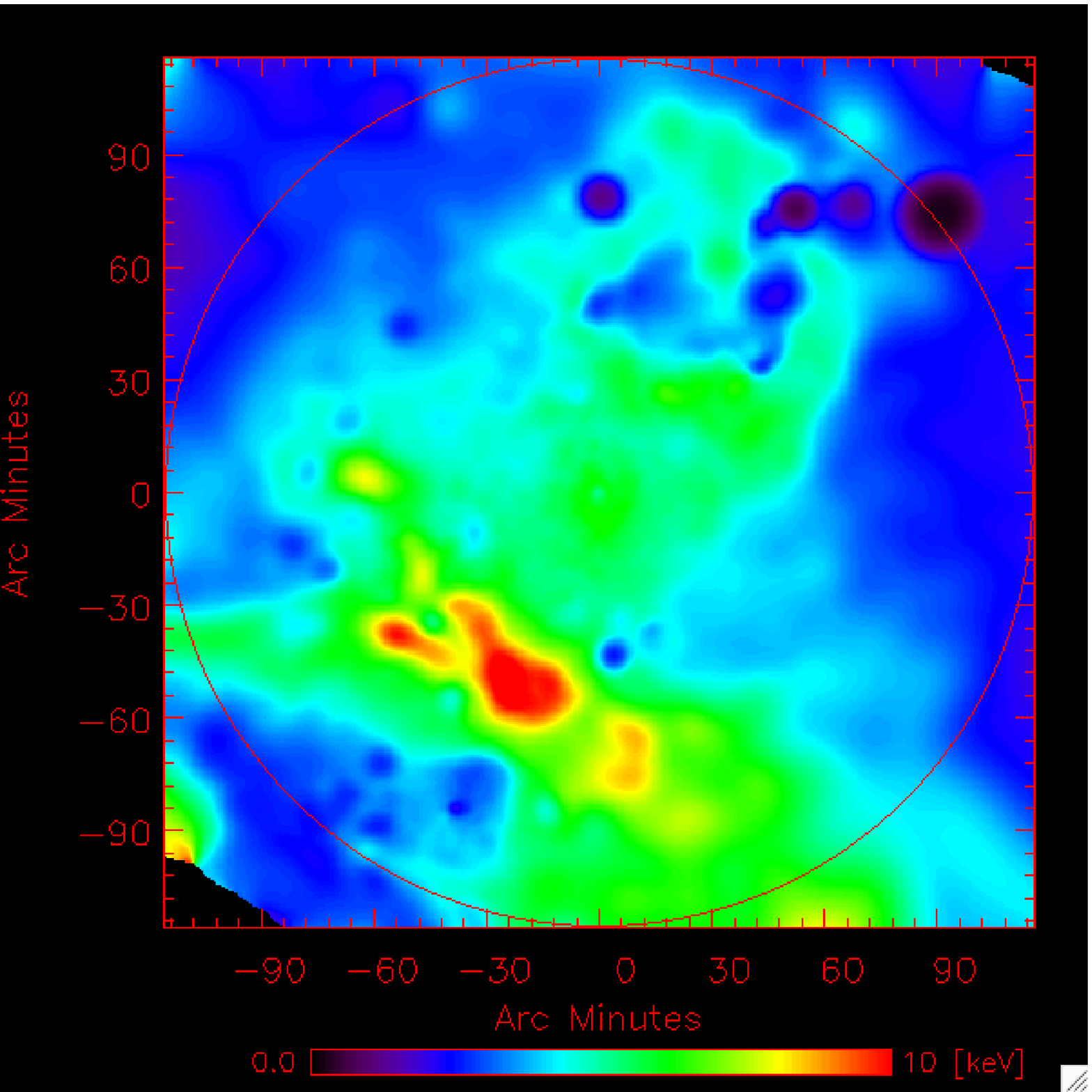}%
\includegraphics*[width=0.45\textwidth]{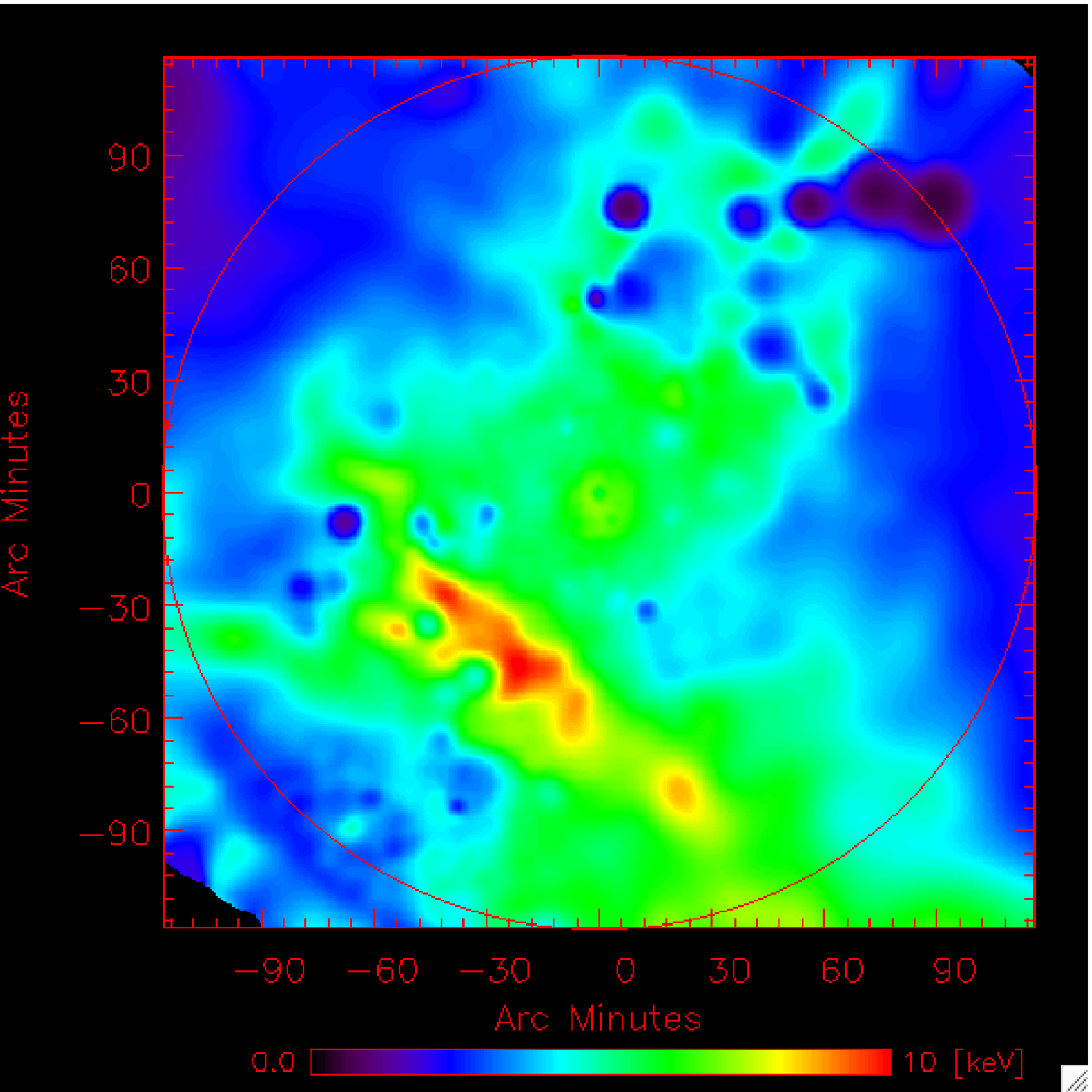}%
\end{center}
\caption{Projected density (top pane) and temperature (bottom) maps
  for one of the most massive clusters found in the simulations. The
  radiative case is presented on the left hand side, whilst a model
  implementing radiative+dust cooling is presented on the right hand
  side {of the figure. Density and temperature maps are
    respectively in logarithmic and linear scales. The overlaid
    circle's radius is the virial radius of the cluster. The X and Y
    axis are given in arcmin considering a projection of the clusrter
    positionned at $z=0.03$ (in this case the spatial scales are just
    indicative, but meaningless).}}

\end{figure}
\subsection{Phase space and cluster maps}
~\\[-3em]
%
On Fig. 2 we present the phase space diagram for the whole box at $z =
0$ for the radiative cooling (left panel) and the radiative-dust
cooling model with $a = 0.001\mu$m and $f_d = 0.1$ (right panel --
model D1 in Table~1).

The displayed quantity is the probability distribution function of gas
particles (computed as the number of particles in logarithmic bins of
size 0.1~dex divided by the total number of particles) as a function
of temperature and overdensity. The colour scale is linear, as
represented on the top of the figure. The features in the diagrams
reflect the gas physical effects included in the simulations:
gravitational heating, adiabatic expansion, radiative cooling and
cooling by dust (the latter effect is only present on the right
panel). Note that our simulations do not include energy feedback nor
photoinization-heating due to a UV background. The absence of the
later effect causes the low density, low temperature, gas to cool
adiabatically without being regulated by photoionization heating. For
$T>10^4$~K gas cools more efficiently according to the cooling rates
of Fig.~1. At high overdensities $\log (\rho/<\rho >) \gtrsim 2.5$ gas
is mainly distributed in bound objects such as galaxy groups and
clusters with high temperatures. The gas in these regions cools
rapidly due to the absence of energy feedback. The inclined V-shape
(in black) on the far right end side of the diagrams is a region of
fast cooling, where the gas cannot remain over a Hubble time. In both
panels, the two upper peaks at intermediate overdensities reflect the
shape of the cooling functions in Fig.~1. The contribution of cooling
due to metals ($Z/Z_\odot =0.3$ at $z=0$ in our simulations) increases
the cooling efficiency at $T=10^6$~K as compared to a gas with
primordial composition. The absence of non-gravitational heating in
our simulations also enhances the appearance of these two upper peaks
in the phase space diagram. The horizontal peak at $T\simeq 10^4$~K
arises because in our simulations the gas below this temperature can
only cool adiabatically. At high overdensities, gas that cools below
$10^4$~K is considered collisionless baryonic matter and it is
converted into galaxy fragments/star forming material.

The differences between the two diagrams are not striking. However,
the concentration of gas in the cool and low-density phase of the D1
model is higher than in the purely gas cooling case (model C). For the
ICM gas, the phase space distributions are also different when looked
in detail.  More particles are present in the form of cold gas (i.e
bellow the horizontal yellow lines) in the radiative+dust cooling
case, as a direct consequence of the extra cooling from the dust
component. As a consequence, the hot and less dense gas distribution
(typically located in the clusters core, i.e. top-right parts of the
diagrams) is more spread towards slightly less hot and dense gas in
the radiative+dust cooling case. This is due to a well-known effect
related with the condensation of gas that is converted into
collisionless star forming material in the simulations. The removal of
cold gas from the hot phase simply leads to an increase of the mean
local temperature of the remaining gas.

%
For the same models we display in Fig.~3 the projected density and
temperature maps for one of the most massive cluster found in our
simulations. As for the phase space, the changes in features on the
density maps are not obvious. However, the locus of high density seems
brighter in the radiative+dust cooling case, e.g the cluster centre is
brighter and more peaked.  The temperature maps present more visible
differences between the dust/no-dust case. Some hot areas are enhanced
or have appeared when adding dust cooling.  This is also a consequence
of the aforementioned (counter intuitive) effect of additional cooling
(e.g due to dust) actually contributes to raise the temperature of the
remaining ICM gas.

\begin{table}
\caption{Parameters of the simulated models  presented on Fig~4.}
\begin{tabular}{lccc}
\hline \hline
Run   & Physics             & $f_d$ & Grain size\\ 
\hline \hline
C & cooling (no dust)   & -            & -      \\
\hline
D1 & cooling with dust   & 0.100        & small  \\
D2 & cooling with dust   & 0.100        & MRN    \\
D4 & cooling with dust   & 0.010        & MRN    \\
\hline \hline
\end{tabular}
\label{tab:mod}
\end{table}

\section{Cluster scaling relations}
\subsection{$T_X-M$ and $L_X-M$ at redshift zero}
~\\[-3em] 
\indent In fact, to be able to properly quantify the effect of the
cooling by dust in the ICM, we have investigated the statistical
properties of the cluster populations forming in our simulations. In each
simulation we selected halos with a total mass larger than $5\times
10^{13}$~M$\odot$. We ended up with a minimum of 60
clusters in each simulations at $z=0$.
Here we illustrate our study with two scaling relations given at
$z=0$: the (emission weighted) temperature--mass relation, $T_X-M$,
and the X-ray bolometric luminosity--mass relation, $L_X-M$. The two
relations are presented on Fig.~4 for the radiative cooling case, and
three radiative+dust models described in Table~1. 

\begin{figure*}
\label{fig:sl}
\hspace*{-0.8cm}\includegraphics*[width=7.5cm]{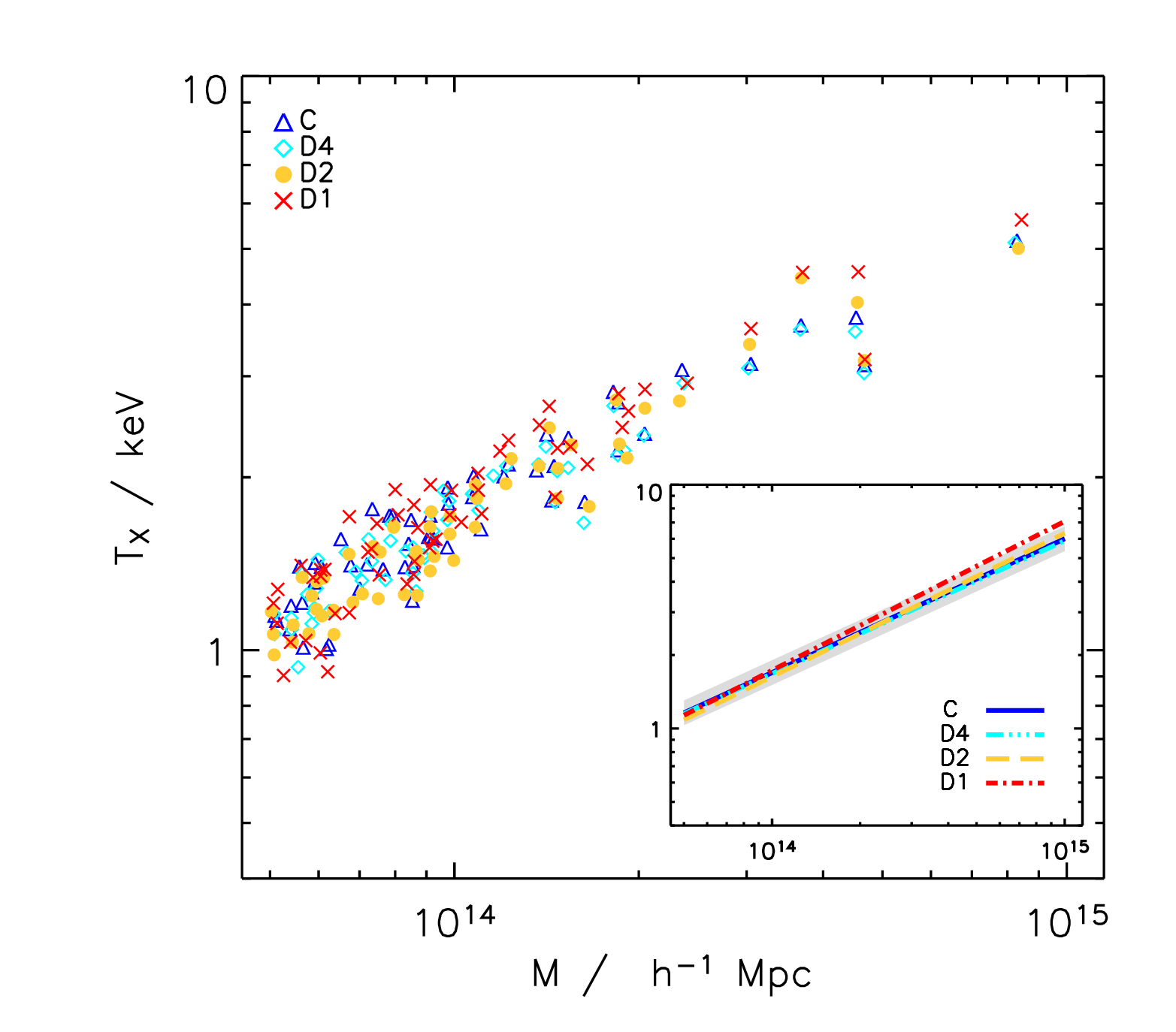}
\includegraphics*[width=7.5cm]{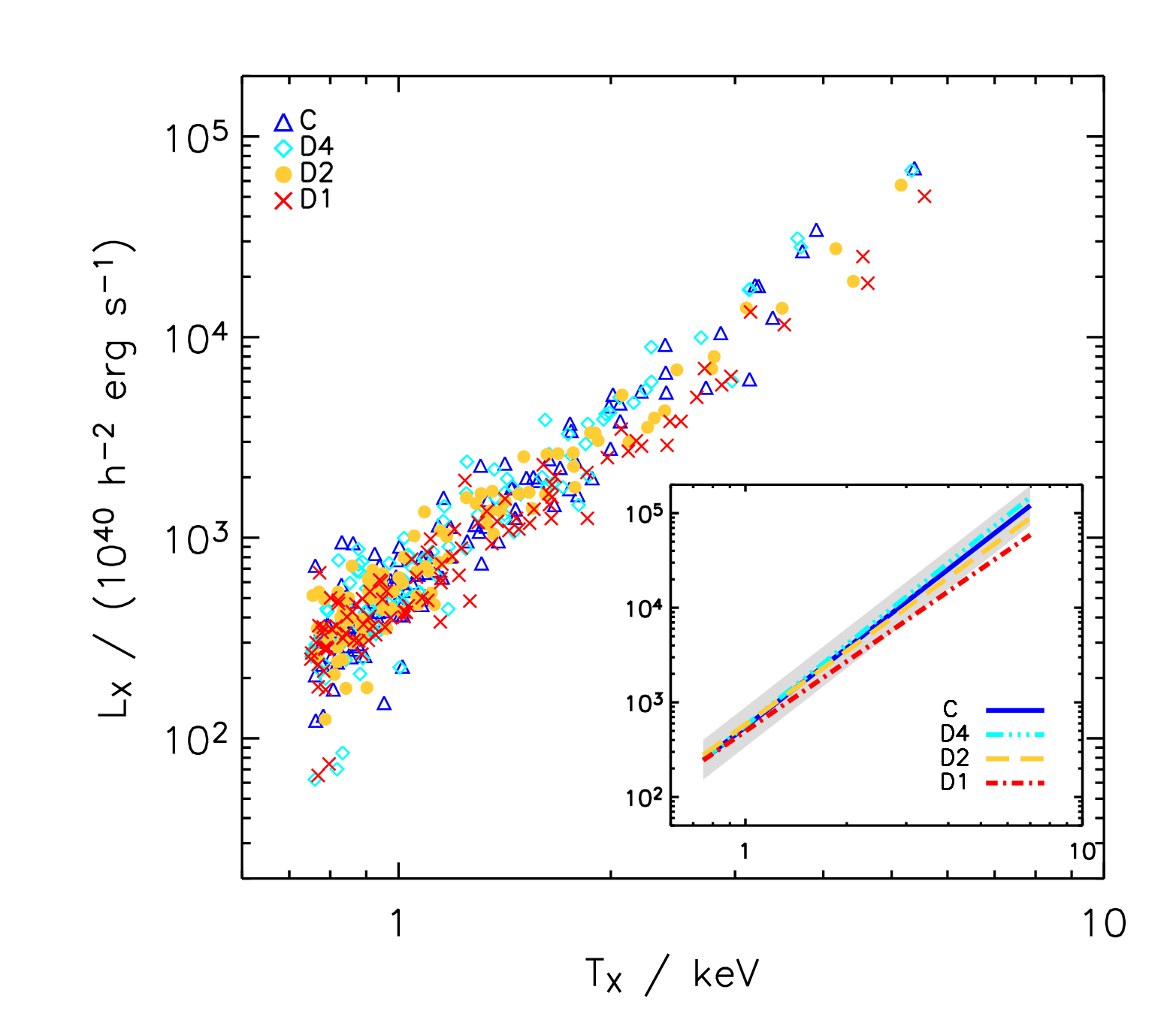}
\caption{(left) Emission weighted temperature--mass relation, $T_X-M$,
  (right) X-ray bolometric luminosity--mass relation, $L_X-M$, at
  redshift zero. Symbols and best fit lines are color-coded for each
  presented model (see Table~1). The grey shaded area corresponds to
  the $\pm 1\sigma$ dispersion for the radiative model (i.e model C).}
\end{figure*}
\indent The $T_X-M$ relation is the less disturbed of the
two. Results for model D4 is almost compatible with the radiative
case. Model D2 and D1 illustrate the effects of the dust
grain size distribution and the dust abundance. 

These small variations are in fact understandable because the
temperature of clusters is essentially dominated by gravity and
because the total cluster mass (that is dominated by the dark mater
component) is in practice unaffected by the extra condensation of a
fraction of the cluster gas mass. However small, we note that the
emission-weighted vs mass scaling is more sensitive to dust than the
mass-weighted vs mass relation studied in \citet{dasilva08}. This
confirms expectations, because $T_X$ gives more weighting to gas with
stronger X-ray emission, i.e gas with high density and temperature
locate in the cluster central regions. As a consequence the $T_X-M$
scaling not only exhibits larger changes of amplitude but also larger
differences in the slopes because of the relative impact of dust on
$T_X$ for different mass systems.  Altogether the variation in
normalisation are less than $\simeq 2$\% over the 3 presented models
at $M=10^{14}$~M$_\odot$.  ~\\[-2em]

On the other hand the $L_X-M$ relation is typically a baryon driven
scaling law. Which means that it is very sensitive to the physics of
baryons at play in the ICM. Again, going from model D4 to D1 via D2,
one can witness the effect of the grain size (as for D1 and D2 the
abundances are the same), and then the abundance effect from model D2
to D4. Both the abundance and the grain size distribution have similar
amplitude, and the overall changes in normalisation at 1~keV goes up
by $\sim 10$\%.  It is also striking how the dust cooling efficiency
increases with the ICM temperature. Taking into account the range of
investigated temperatures, this is perfectly consistent with the dust
cooling function implemented in the simulations, thus with the study
of \citet{montier04} that shows this discriminating effect with the
temperature.  ~\\[-2em]

Also, we have pointed out that these changes induced by dust cooling
on the normalisation as well as the slope of the scaling relations,
are strongly linked to the physical properties of dust. We have
illustrated that the effect of the dust grain size is as efficient as
the effect of the abundance of dust grains populating the ICM. ~\\[-2em]

\subsection{Limitations of the model}
~\\[-3em] 
%
It is obvious that our present implementation of dust cooling in
simulations has intrinsic limitations and should be considered as a
first (``zero order'') approach to the problem, see
\citet{dasilva08}. Indeed, with our implentation of dust in our
simulations, we directly linked the presence (thus abundance) of dust
with the presnce of metals (i.e we put dust wherever there are metals
in the simulations). However, dust and metals are not sensitive to the
same physical processes, and if the metals are not destroyed, the dust
grains are sputtered. From \citet{draine79} and \citet{montier04}, for
a hot thermal plasma, i.e $10^6<T<10^9$~K, and densities met in the core
of dense clusters, i.e $n_H\sim 10^{-3}$~cm$^{-3}$, and grain sizes
ranging from $0.001$ and $0.5\; \mu$m, dust grains have lifetimes of
$10^6<t_{dust}<10^9$~yr. The overall efficiency of cooling by dust in
the ICM/IGM is thus strongly linked to the physical processes of dust
injection.  Therefore, beside the cooling function of dust, with which
we have performed a fully self-consistent implementation of the effect
of dust as a cooling vector of the ICM/IGM, we did not performed {\it
  stricto senso} a physical implementation, as we do not deal with the
physics of the dust creation and dust destruction processes.  This
would be the next step in our study.  ~\\[-2em]

Also, we directly correlated the abundance of dust with
metallicity. The evolution law we chose for the metallicity does
underestimate the metallicity at high redshift. Indeed,
$Z=0.3(t/t_0)Z_\odot$ normalised at $0.3Z_\odot$ at $z=0$ gives $\sim
0.2$ at $z=0.5$ and $\sim 0.1$ at $z=1$. However, observations as well
as theoretical studies on the metal enrichment of the ICM have
indicate high values of the metalicity, i.e $Z\sim 0.3 Z_{\odot}$, up
to redshift above 1 \citep{cora08,borgani08}. An early enrichment of
the ICM/IGM with dust, then sputtered, could (partly) explain these
high values of the metallicity. Therefore, by underestimating the
metallicity at high redshifts, we may have underestimated the amount
of dust injected in the ICM at high redshifts, and thus the efficiency
of dust cooling when integrated from an early epoch down to redshift
zero ~\\[-2em]

{They are probably other limitations to our results, such as the local
  thermal equilibrium hypothesis underlied in the computation of
  \citet{montier04} cooling function (following a non-equilibrium
  hypothesis \citep{dwek90}), that may lead to difference in the cooling
  curves, the used theoretical setup for
  the dust emission within an optically thin hot thermal plasma is
  well suited to our purpose of assessing the influence of dust on the
  galaxy clusters properties.}  ~\\

\section{Conclusion}
With our current ``zero'' order implementation of
the effect of dust cooling, we have demonstrated the actual effect of
the presence of dust within the IGM/ICM as a non-gravitationnal
process that plays a role on the physical properties of baryons within
large scale structures.  Deeper studies including other scaling laws
such as the entropy-mass, $S-M$, and the integrated Comptonisation
parameter-mass, $Y-M$, relations are presented in \citet{dasilva08}.


~\\ {\it Acknowledgements: We thanks the organisers of the session E14
  ``The interplay between the Interstellar and Intergalactic Media
  from High Redshifts to Present'' of the 37$^{\textrm{th}}$ COSPAR
  assembly, D. Wang and M. Shull, during which the this work was
  presented. We are also grateful to the two anonymous referees for
  their fruitful commnents on our manuscript. MR, LM and EP are
  supported by grant ANR-06-JCJC-0141. AdS was supported by Funda\c
  c\~ao Ci\^encia Tecnologia (FCT) Portugal under contracts
  SFRH/BPD/20583/2004 and CIENCIA 2007.}

\end{document}